\documentclass[%
 reprint,
superscriptaddress,
 amsmath,amssymb,
 aps,
]{revtex4-1}

\usepackage{graphicx}
\usepackage{dcolumn}
\usepackage{bm}
\usepackage{color}
\newcommand{\mbf}[1]{\mathbf{#1}}

\begin{document}

\preprint{APS/123-QED}

\title{Metamaterials for Active Colloid Transport}

\author{Shahrzad Yazdi}
\affiliation{Department of Materials Science and Engineering, Massachusetts Institute of Technology, Cambridge, MA, 02139, USA}
\author{Juan L. Aragones} 
\affiliation{Department of Materials Science and Engineering, Massachusetts Institute of Technology, Cambridge, MA, 02139, USA}
\affiliation{\textit{present address:} IFIMAC, Facultad de Ciencias, Universidad Autonoma de Madrid, Ciudad Universitaria de Cantoblanco,
28049, Madrid, Spain}
\author{Jennifer Coulter}
\affiliation{Department of Materials Science and Engineering, Massachusetts Institute of Technology, Cambridge, MA, 02139, USA}
\affiliation{\textit{present address:} John A. Paulson School of Engineering and Applied Sciences, Harvard University, Cambridge, MA, 02138, USA}
\author{Alfredo Alexander-Katz}
\email{aalexand@mit.edu}
\affiliation{Department of Materials Science and Engineering, Massachusetts Institute of Technology, Cambridge, MA, 02139, USA}



\begin{abstract}
Transport phenomena in out-of-equilibrium systems is immensely important in a myriad of applications in biology, engineering and physics. Complex environments, such as the cytoplasm or porous media, can substantially affect the transport properties of such systems. In particular, recent interest has focused on how such environments affect the motion of active systems, such as colloids and organisms propelled by directional driving forces. Nevertheless, the transport of active matter with non-directional (rotational) activity is yet to be understood, despite the ubiquity of rotating modes of motion in synthetic and natural systems. Here, we report on the discovery of spatiotemporal metamaterial systems that are able to dictate the transport of spinning colloids in exquisite ways based on solely two parameters: frequency of spin modulation in time and the symmetry of the metamaterial. We demonstrate that dynamic modulations of the amplitude of spin on a colloid in lattices with rotational symmetry give rise to non-equilibrium ballistic transport bands, reminiscent of those in Floquet-Bloch systems. By coupling these temporal modulations with additional symmetry breaking in the lattice, we show selective control from 4-way to 2-way to unidirectional motion. Our results provide critical new insights into the motion of spinning matter in complex (biological) systems. Furthermore, our work can also be used for designing systems with novel and unique transport properties for application in, for example, smart channel-less microfluidics, micro-robotics, or colloidal separations.
\end{abstract}

\maketitle

Active transport of particles or organisms has been a topic of interest in multiple fields ranging from molecular transport of proteins \cite{vale2003molecular} to the motion of phytoplankton \cite{durham2009disruption} to the self-organization of flocking birds \cite{ballerini2008interaction}. At small scales, inhomogeneous environments strongly affect the transport of self-propelled/driven particles \cite{bechinger2016active, morin2017distortion} and microorganisms \cite{lauga2009hydrodynamics, molaei2014failed}. Surprisingly, self-propelled organisms dynamically alter their propulsion mechanism as they receive hydrodynamic feedback from obstacles in their surrounding environment. For example, the nematode \textit{C. elegans} swims faster in porous environments by coupling its undulatory locomotion mechanism to match the periodicity of the pillars that make the porous media \cite{majmudar2012experiments}. In another study, Engstler and co-workers have shown how adaptation of locomotion mechanisms in certain parasites to the density of neighboring obstacles helps them endure harsh crowded environments like blood \cite{heddergott2012trypanosome}. These studies showcase that boundary conditions strongly affect the motion of organisms in heterogeneous media at microscale. However, what remains to be understood is how the interplay between spatial arrangement of the heterogeneities in the environment with the characteristics of the propulsion mechanism can lead to a wide variety of non-trivial transport regimes, such as trapped or ballistic? 

Inspired by these natural systems, we approach this problem from a different point of view and ask: for a given mode of activity, can one design materials that precisely control the transport of active particles? Here, we show that this is possible for a class of active colloids that spin instead of self-propel. Spinning colloids, while non-motile on their own, have been shown to display new emergent states different from those of self-propelled particles \cite{van2016spatiotemporal,nguyen2014emergent,aragones2016elasticity,steimel2016emergent}. In particular, for active spinning colloids, the environment plays a critical role since they do not exhibit translational motion in a homogeneous environment, unless they are close to boundaries or other particles \cite{sing2010controlled}. In order to completely define the transport of active spinning colloids, we incorporate insights from out-of-equilibrium quantum systems, and in particular from Floquet-Bloch systems \cite{zhang2017observation,choi2017observation,dal2015floquet}, where the combination of temporal modulation of the potential and the spatial arrangement of the lattices can lead to non-trivial transport states \cite{lohse2016thouless,mahmood2016selective,fujiwara2019transport,esin2017steady}. In our particular colloidal system, we couple the temporal modulation of the activity to the symmetry of the underlying substrate to create metamaterials that control the transport of active colloids. These metamaterials exhibit controlled transport in any direction of the lattice and can be predefined by appropriate selection of the symmetry of the environment together with the frequency of modulation, as will be shown below. We believe our proposed model could be helpful in designing smart materials for controlled transport and separation of colloidal particles based on their physicochemical properties which is important for many fields including microfluidics and soft robotics. Also, the physical principles unraveled in our active colloidal system may pave the way to a predictive understanding of non-thermal transport in active biological systems.
\begin{figure}[ht]
\centerline{\includegraphics[clip,scale=0.21,angle=-0]{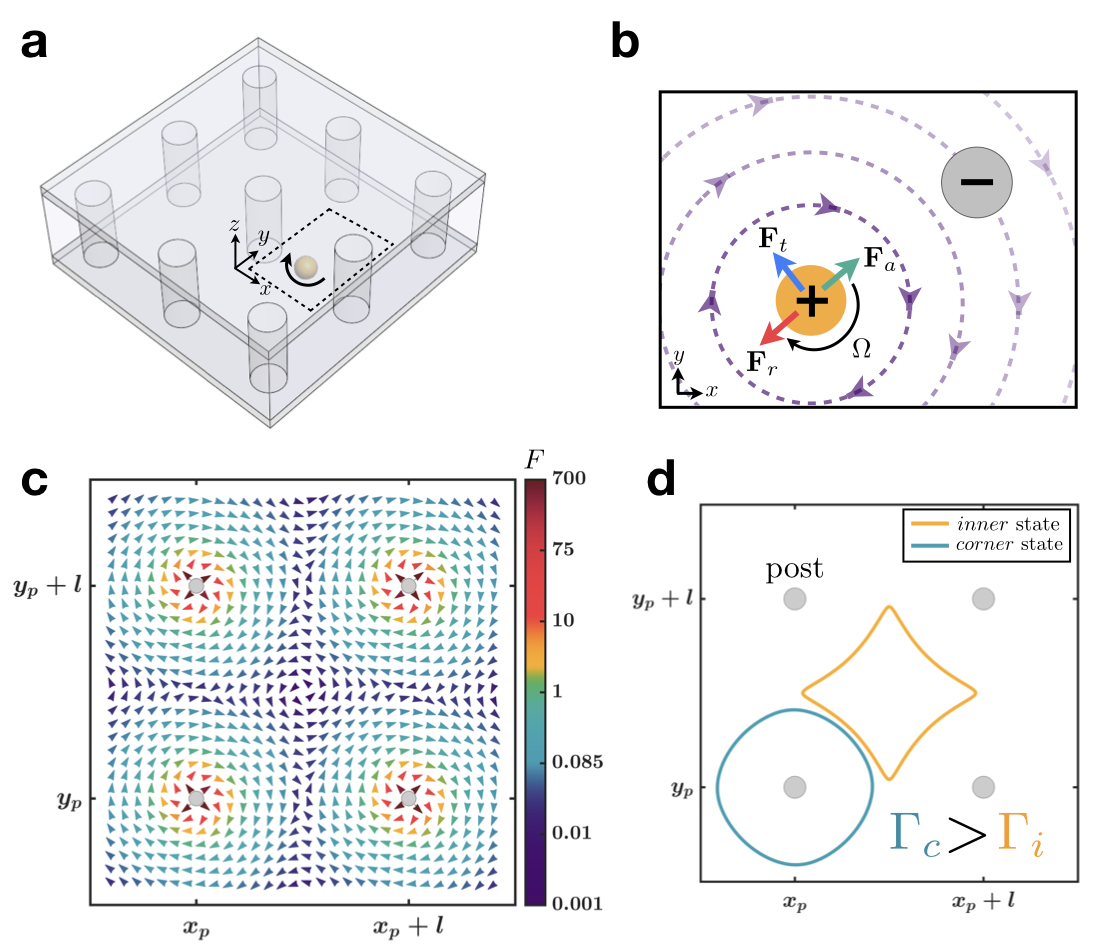}}
\caption{Active particle spinning in a periodic array of posts. \textbf{a} Schematic of the system geometry: a colloid spins in a 2D array of posts in a Newtonian fluid. \textbf{b} Forces on a spinner with angular velocity $\boldsymbol{\Omega}$ in the lattice are: the hydrodynamic repulsive and rotational (tangential) forces, $\mbf{F}_r$ and $\mbf{F}_t$ respectively. The attractive force  $\mbf{F}_a$ that is due to electrostatic interactions. \textbf{c} An example of the effective force field is shown in the unit cell of a square lattice. The color indicates the magnitude of the net force. The conditions used in this plot are : $l = 5$, $Re^t = 0.17$, A $= 0.05$, B $= 1$ , $\Gamma = 0$ and no thermal fluctuations, i.e. Pe $\rightarrow \infty$). \textbf{d} The steady state solution of this system in the absence of activity modulation exhibits two states: \textit{inner} band (yellow) with $\Gamma_i =  5.2 \times 10^{-4}$ and \textit{corner} band (blue) with $\Gamma_c = 0.01$. Due to the rotational symmetry of the square lattice, both states of the spinner are localized.  \label{fig1}}
\end{figure}  

Our system is composed of an active particle and a passive 2D array of fixed posts in a Newtonian fluid of viscosity $\eta$ confined between two substrates (Fig.~\ref{fig1}~a). The active particle is spinning with its axis of rotation perpendicular to the confining substrates ($z$-axis). As the spinning particle (or spinner) rotates, it creates a vortical velocity field. In the limit of small but finite Reynolds number, this field contains a purely rotational component, $\mbf{F}_t$, and a small, albeit important, radial component, $\mbf{F}_r$. (Fig.~\ref{fig1}~b). The rotational (tangential) component causes the colloid to rotate around the posts and it scales as $1/r^3$, where $r$ is the distance of the spinner from the post. The secondary flows induce a radial repulsive force $\mbf{F}_r$ between the spinner and the posts that has a more complex scaling, and was determined using Lattice Boltzmann simulations, which show excellent agreement with experiments \cite{aragones2016elasticity,steimel2016emergent}. Finally, the attraction $\mbf{F}_a$ between the colloid and the posts is considered for simplicity as a simple Coulomb force. We note that our conclusions are general even if we replace this force with a screened interaction or another attraction force of any nature (for example depletion forces). A typical force field experienced by the spinner in this environment is shown in Fig.~\ref{fig1}~c. For convenience, the force is made dimensionless using characteristic Stokes drag force $6 \pi \eta \Omega_0 r_s^2$, where $r_s$ is the radius of the colloid and $\Omega_0$ is the constant rotational frequency of the spinner. The force on the spinner in the dimensionless form is written as 
\begin{equation}
\mbf{F} = \sum_i \text{A}~Re~\hat{\mbf{U}} \times \hat{\boldsymbol{\Omega}}/r^3_i ~ S(Re)+ \text{B}~\hat{\boldsymbol{\Omega}} \times \hat{\mbf{r}}_i /r^3_i - \Gamma~\hat{\mbf{r}}_i/r_i^2,~ \label{eq1}
\end{equation}
where the overall magnitude of the force is dictated by the parameters A, B, and $\Gamma$ that corresponds to the rotational speed, physical and electrostatic properties of the medium, posts, and spinner. The first two terms correspond to the hydrodynamic repulsion ($\mbf{F}_r$) and rotational ($\mbf{F}_t$) forces, respectively. The repulsive hydrodynamic term is a lift force on the spinner from the neighboring posts at low, but finite Reynolds number, $Re = \rho r_s^2 \Omega_0/\eta$, where $\rho$ is the density of the fluid. According to our numerical solution of the Navier-Stokes equations using the Lattice Boltzmann method, this Magnus lift force is only effective beyond a certain \textit{transition} Reynolds number, $Re^t$. When the spinner's motion corresponds to $Re < Re^t$, the magnitude of this lift  force becomes negligible compared to the other forces (see SI). We implemented this behavior in our Langevin model using a sigmoid function, $S(Re) = 1/(1 + e^{-G(Re - Re^t)})$, where $G$ determines the width of the transition region. Electrostatic forces are described by the third term in Eq.~\ref{eq1}. In this equation, the sum runs in principle through all the obstacles indexed by the parameter $i$. The distance between the spinning particle and the $i$th obstacle is given by $r_i$ and the hat denotes unitary vectors.  

\begin{figure}[ht]
\centerline{\includegraphics[clip,scale=0.26,angle=-0]{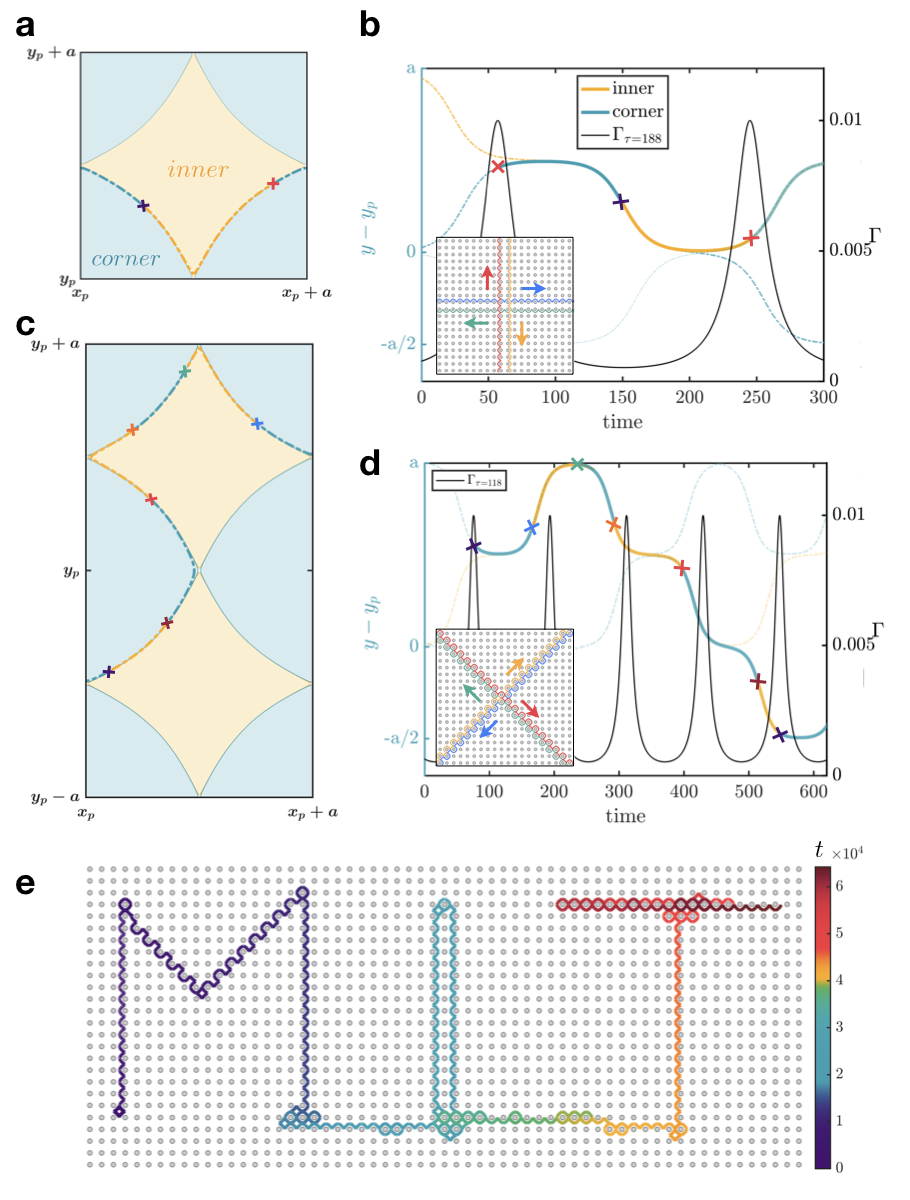}}
\caption{Ballistic transport in a square lattice can be induced via dynamic hybridization of the \textit{inner} and \textit{corner} localized states by modulating the rotational frequency of the spinner. \textbf{a} and \textbf{c} Domains of localized solutions in a square unit cell are shown in yellow and blue. For $\Gamma_0 < 0.0021$, the spinner is localized at \textit{inner} domain plotted in yellow and at $\Gamma_0 \geq 0.0021$, it is bound to the blue \textit{corner} domain. \textbf{b} and \textbf{d} Modulating the angular velocity $\Omega(t)$ at time periods of $\tau = 188$ and $\tau = 118$, leads to two 2D modes of ballistic motion. Insets of \textbf{b} and \textbf{d} show \textit{straight} and \textbf{f} \textit{diagonal}, trajectories, respectively. \textbf{e} As a demonstration, we wrote MIT letters in a continuous trajectory by switching between these two time periods ($\tau = 118$ and $188$) in a timely manner. The parameters are the same as Fig.~\ref{fig1} except for $\Gamma_0 = 0.001$ and $a_0 = 0.9$.  \label{fig2}}
\end{figure}

To understand under what conditions this system allows ballistic transport modes, we directly simulate the dynamics of the spinning colloid in the overdamped regime using the dimensionless Langevin equation  
\begin{equation}
\dot{\mbf{r}} = \mbf{F} + \sqrt{\frac{2}{Pe}} \ \boldsymbol{\xi}(t) \label{eq2}
\end{equation}
where $\mbf{r}$ is the position and $ \boldsymbol{\xi}(t)$ is a normalized white noise with $\langle \xi(t) \rangle = 0$ and $\langle \xi(t) \xi(t') \rangle = \delta(t - t')$. The deterministic forces ($\mbf{F}$), including both hydrodynamic and non-hydrodynamic forces, are obtained from Eq.~\ref{eq1}. The P{\' e}clet number (Pe) measures the ratio of convective to thermal (stochastic) forces and it is defined as Pe  $= \Omega_0 r_s^2/D$, where $D = \frac{k_BT}{6\pi\eta r_s}$ is the diffusion coefficient. Eq.~\ref{eq2} is solved using an explicit Euler forward time scheme. The equivalent Fokker-Planck equation for this system corresponds to a Convection-Diffusion equation, where the repulsive term in the force becomes the reaction/potential term, and the rotational component of the force becomes the convective term. The diffusive part corresponds to the random motion of the spinner characterized by Pe. Our design for this material was inspired by the several similarities that the Fokker-Planck equation shares with the Schr$\ddot{\text{o}}$dinger equation for 2D electrons in magnetic fields in a periodically pinned system. 

In the case of a 2D square array, and in the absence of any modulation of the rotational speed (i.e. constant $\Omega = \Omega_0$), the spinner is localized in one of two states (Fig.\ref{fig1}~d) : i) \textit{inner} state and ii) \textit{corner} state. Both of these states are localized and display no net motion, differing only by a post ``winding number'': 0 for \textit{inner} state and 1 for \textit{corner} state.

Inspired by Floquet-Bloch bands in optical lattices, where one can control transport of quantum systems \cite{fujiwara2019transport}, we modulate the magnitude of spinner's rotational speed as
\begin{equation}
\Omega(t) = \Omega_0\big(1 + a_0\sin(\frac{2\pi}{\tau} t)\big), \label{eq3}
\end{equation}
in order to controllably hybridize both bound (localized) states observed in the non-modulated case. Here, $t$ is time and $\tau$ is the period of modulation. Note that in our dimensionless formulation shown in Eq.~\ref{eq1}, this modulation is applied via $\Gamma = \Gamma_0/(1 + a_0\sin(\frac{2\pi}{\tau} t))$. As shown in Fig.~\ref{fig2}, an amplitude-modulated spinner in a square lattice can cross between the \textit{corner} and \textit{inner} trajectories (or bands) and eventually break free of these ``localized'' states and transport ballistically between unit cells. We present two hybridization patterns of these bands. We start by finding the $\Gamma_0$ parameter that corresponds to steady solution of the spinner at the most outer \textit{corner} band. Application of an attraction force with such magnitude assures that the spinner, regardless of its initial position, moves to a trajectory at the edge of \textit{corner} and \textit{inner} states (Fig.~\ref{fig2}~a and c). If we keep $\Omega$ constant, the spinner remains on the most outer \textit{corner} band and follows a closed loop around the neighboring post, as shown with a blue-dotted line in Fig.~\ref{fig2}~a and c. However, if we periodically oscillate $\Omega$ (i.e. $\Gamma$) such that the spinner moves to the \textit{inner} band when $\Gamma$ is minimum and transfers to the \textit{corner} bands at $\Gamma$'s peak, we can transport the spinner between unit cells. The time period at which this $\Gamma$ modulation should occur is dictated by the speed of the spinner on these two closed bands and the number of times the spinner crosses between localized states (Fig.~\ref{fig2}~b and d). For the particular set of parameters we have chosen here ($l = 5$, $Re^t = 0.17$, A $= 0.05$, B $= 1$ , $\Gamma_0 = 0.001$, $a_0 = 0.9$, and no thermal fluctuations, i.e. Pe $\rightarrow \infty$), we find that at $\tau = 188$ the spinner crosses between the bands only twice in each unit cell, which yields a 4-fold degenerate \textit{straight} ballistic trajectory (Fig.~\ref{fig2}~b inset) along the (1 0) or (0 1) planes of the array. Alternatively, for higher number of crossings in a unit cell, the spinner can move ballistically on a \textit{diagonal} path, which for our system occurs at $\tau = 118$ (Fig.~\ref{fig2}~d inset). Finally, by dynamically adjusting $\tau$ one can control the transport of the colloid in any desired pattern (Fig.~\ref{fig2}~e). More discussion on the ranges of $\tau$ and the kinetics of transitions between these states are available in the SI. Furthermore, to gain a deeper understanding of the conditions under which ballistic transport is achieved in square lattices, we probed the design parameters in the Floquet-Bloch system (see SI). For this system, we found a quadratic relation between angular modulation frequency $2\pi/\tau$ and the lattice wavenumber $2\pi/\lambda$ (see the quasi-dispersion plot in Fig.~S6), reminiscent of that for a free electron.  

\begin{figure}[ht]
\centerline{\includegraphics[clip,scale=0.25,angle=-0]{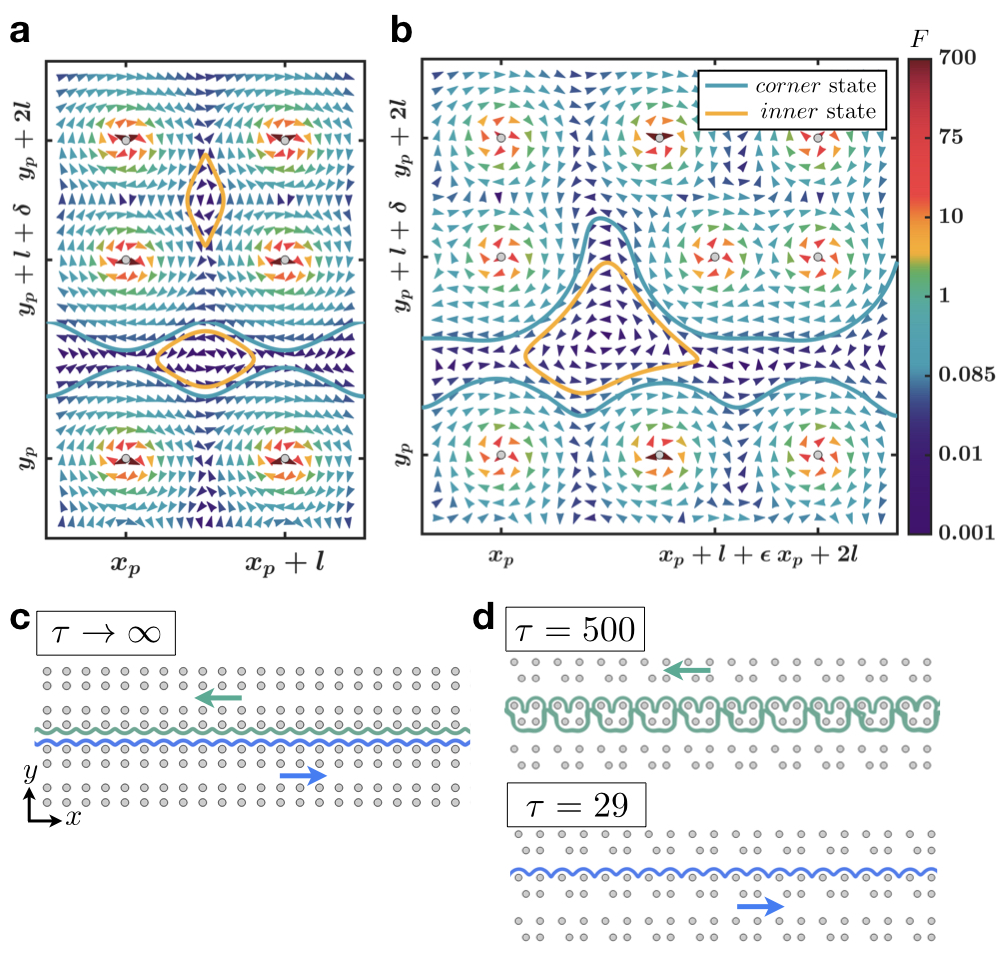}}
\caption{Breaking the rotational symmetry in the lattice leads to emergence of ballistic bands, even in the absence of amplitude modulation. The effective force field of Y-Shifted and XY-Shifted lattices are shown in \textbf{a} and \textbf{b}, respectively. The parameters are as stated in Fig.~\ref{fig1}b. The broken rotational symmetry leads to the opening of \textit{corner} states into ``directionality avenues'' in the $y$-direction, which yields to the \textbf{c} emergence of ballistic motion of the spinner with no modulation in amplitude of $\Omega$. \textbf{d} When modulating $\Omega$ in the XY-Shifted lattice one can control the directionality of the spinner. For example, all the spinners move in the positive (negative) direction of $x$-axis at $\tau = 29$ ($\tau = 500$). The parameters are the same as in Fig.~\ref{fig1} except for $\Gamma_0 = 0.003$ and $a_0 = 0.95$. Results here are shown for $\delta = 1.25$ and $\epsilon = 1.75$. \label{fig4}}
\end{figure}

In the previous part, we found 4-way transport regimes along the square lattice symmetry planes. However, one can reduce the degeneracy by breaking the symmetry of the system. Two examples of lattices with broken rotational symmetries are shown in Fig.~\ref{fig4}~a and b. The rotational symmetry in the lattice in Fig.~\ref{fig4}~a is broken by shifting every other row of the posts in a square lattice by $\delta y = \delta$, henceforth referred to as ``Y-Shifted'' lattice. The lattice in Fig.~\ref{fig4}~b has also a broken inversion symmetry, since the position of every other post in the shifted rows, is moved by the amount $\delta x = \epsilon$, hence called ``XY-Shifted''. 

As shown in Fig.\ref{fig4}~a and b, the broken rotational symmetry in both Y-Shifted and XY-Shifted lattices alter \textit{corner} states to open trajectories. Note that the addition of the attraction force on the spinner can eliminate the \textit{inner} states by merging them with open \textit{corner} trajectories. There are two main differences between the ballistic transport states of the spinner in these lattices with broken rotational symmetries and that in square lattices. First, the symmetry of the array of posts in Y-Shifted and XY-Shifted lattices is sufficient for a ballistic state to emerge, hence, there is no need for amplitude-modulation to hybridize the closed bands. This symmetry-protected transport displays significant robustness against thermal fluctuations (see SI) compared to the transport in square lattices driven by amplitude modulation. Second, the degree of freedom of the motion vanishes in the direction that the rotational symmetry is broken (i.e. $y$-direction). Therefore, the spinner is only restricted to move in one-dimension (Fig.~\ref{fig4}~c and d). Nevertheless, we show that by breaking the inversion symmetry (as in XY-Shifted lattice), the amplitude-modulated $\Omega$ can be used to further reduce the motion's degree of freedom to either positive or negative $x$-direction, depending on $\tau$ (Fig.~\ref{fig4}~d). These results show that a Floquet-Bloch system with broken symmetries in the lattice can be used to design metamaterials capable of transporting active spinning particles in unidirectional fashion. While we applied this concept in the $x$-direction, in general this will apply in the $y$-direction, and thus one can create completely controlled transport of active colloids by either spatially having different lattices or by using a single lattice and modulating the rotational speed $\Omega$.   

In summary, here we have introduced a new approach to design material systems which offer exquisite control of the transport regimes of active colloids. Such metamaterials rely on the selective hybridization of bound states and the coupling between the spatial symmetry of the lattice and the modulation of the activity. While in this work we only focused on spinning particles, we believe our results are equally applicable to other vortical (spinning) flows. Examples include, but are not limited to, emergent of vortices in bacterial suspensions \cite{wioland2016ferromagnetic,nishiguchi2018engineering} or in populations of colloidal rollers \cite{bricard2015emergent,kokot2013emergent,kokot2018manipulation}. Also, our work can be extended to other lattices with higher symmetries where more planes for transport may be available, and thus more control on the directionality of the system.  Nevertheless, our results are general and show that in a symmetric lattice it is possible to hybridize bound states by an amplitude-modulated spin to engineer new transport regimes. By sequentially breaking the rotational symmetry, we explicitly showed that one can open ``directionality avenues'' along which the active particles can spontaneously move ballistically, opening the door for complete control of the motion of a colloid in these lattices. Finally, we note that the ballistic motion of the spinners in these lattices is reminiscent of the motion of electrons on the edge of a 2D electron gas in the presence of strong perpendicular magnetic field, as in the case of the quantum Hall effect. This behavior is also intimately related to the mass transport of active nematic swimmers in Lieb lattices with grain boundaries \cite{souslov2017topological}. Further work is necessary to understand the full prospects of these non-equilibrium states of active spinning systems, but this work is a first incursion into this unexplored area. We believe our results have many potential implications in the way vortices in nature move in different systems and will have direct applications in microfluidics, micro-robotics, and particle separation schemes. 

\section*{Acknowledgments}
S.Y. is grateful for support provided by the Burroughs Wellcome Fund through their Career Award at the Scientific Interface. J.C. and A.A.K. thank the MIT MRSEC Program of the National Science Foundation for funding of the REU program under award DMR 14-19807. J.A. acknowledges the support of a fellowship from ”la Caixa” Foundation (ID 100010434 and code number LCF/BQ/LI18/11630021) and also IFIMAC (MDM-2014-0377). This work was partially supported by the Department of Energy BES Award No. ER46919 (LB simulations).

\bibliography{Refs}

\end{document}